\pdfoutput=1

\documentclass[12pt,a4paper]{article}

\usepackage{ifthen} 
\newboolean{pdflatex}
\setboolean{pdflatex}{true} 

\newboolean{articletitles}
\setboolean{articletitles}{true} 

\newboolean{uprightparticles}
\setboolean{uprightparticles}{false} 

\newboolean{inbibliography}
\setboolean{inbibliography}{false} 

\usepackage{microtype}
\usepackage{lineno}  
\usepackage{xspace} 
\usepackage{caption} 

\usepackage{graphicx}  
\usepackage{color}
\usepackage{colortbl}
\graphicspath{{./figs/}} 

\usepackage{amsmath} 
\usepackage{amssymb}
\usepackage{amsfonts}
\usepackage{upgreek} 

\newcommand*\patchAmsMathEnvironmentForLineno[1]{%
\expandafter\let\csname old#1\expandafter\endcsname\csname #1\endcsname
\expandafter\let\csname oldend#1\expandafter\endcsname\csname
end#1\endcsname
 \renewenvironment{#1}%
   {\linenomath\csname old#1\endcsname}%
   {\csname oldend#1\endcsname\endlinenomath}%
}
\newcommand*\patchBothAmsMathEnvironmentsForLineno[1]{%
  \patchAmsMathEnvironmentForLineno{#1}%
  \patchAmsMathEnvironmentForLineno{#1*}%
}
\AtBeginDocument{%
\patchBothAmsMathEnvironmentsForLineno{equation}%
\patchBothAmsMathEnvironmentsForLineno{align}%
\patchBothAmsMathEnvironmentsForLineno{flalign}%
\patchBothAmsMathEnvironmentsForLineno{alignat}%
\patchBothAmsMathEnvironmentsForLineno{gather}%
\patchBothAmsMathEnvironmentsForLineno{multline}%
\patchBothAmsMathEnvironmentsForLineno{eqnarray}%
}

\usepackage{hyperref}    
\usepackage[all]{hypcap} 

\def\lhcb {\mbox{LHCb}\xspace}

\ifthenelse{\boolean{uprightparticles}}%
{

 \def\Ppi         {\ensuremath{\uppi}\xspace}

 \def\Ppsi        {\ensuremath{\uppsi}\xspace}

 \def\PDelta      {\ensuremath{\Delta}\xspace}                 
 \def\PXi      {\ensuremath{\Xi}\xspace}                 
 \def\PLambda      {\ensuremath{\Lambda}\xspace}                 
 \def\PSigma      {\ensuremath{\Sigma}\xspace}                 
 \def\POmega      {\ensuremath{\Omega}\xspace}                 
 \def\PUpsilon      {\ensuremath{\Upsilon}\xspace}                 
 
 \def\PB      {\ensuremath{\mathrm{B}}\xspace}                 
                  
 \def\PD      {\ensuremath{\mathrm{D}}\xspace}

 \def\PJ      {\ensuremath{\mathrm{J}}\xspace}                 
 \def\PK      {\ensuremath{\mathrm{K}}\xspace}

 \def\Pb      {\ensuremath{\mathrm{b}}\xspace}                 
 \def\Pc      {\ensuremath{\mathrm{c}}\xspace}

 \def\Pi      {\ensuremath{\mathrm{i}}\xspace}

 \def\Ps      {\ensuremath{\mathrm{s}}\xspace}

}
{

 \def\Ppi         {\ensuremath{\pi}\xspace}

 \def\Ppsi        {\ensuremath{\psi}\xspace}                 
                  
 \mathchardef\PDelta="7101
 \mathchardef\PXi="7104
 \mathchardef\PLambda="7103
 \mathchardef\PSigma="7106
 \mathchardef\POmega="710A
 \mathchardef\PUpsilon="7107
                  
 \def\PB      {\ensuremath{B}\xspace}                 
                  
 \def\PD      {\ensuremath{D}\xspace}

 \def\PJ      {\ensuremath{J}\xspace}                 
 \def\PK      {\ensuremath{K}\xspace}

 \def\Pb      {\ensuremath{b}\xspace}                 
 \def\Pc      {\ensuremath{c}\xspace}

 \def\Pi      {\ensuremath{i}\xspace}

 \def\Ps      {\ensuremath{s}\xspace}

}

\makeatletter
\ifcase \@ptsize \relax
  \newcommand{\miniscule}{\@setfontsize\miniscule{4}{5}}
\or
  \newcommand{\miniscule}{\@setfontsize\miniscule{5}{6}}
\or
  \newcommand{\miniscule}{\@setfontsize\miniscule{5}{6}}
\fi
\makeatother

\DeclareRobustCommand{\optbar}[1]{\shortstack{{\miniscule (\rule[.5ex]{1.25em}{.18mm})}
  \\ [-.7ex] $#1$}}

\def\squark    {{\ensuremath{\Ps}}\xspace}

\def\cquark    {{\ensuremath{\Pc}}\xspace}

\def\bquark    {{\ensuremath{\Pb}}\xspace}

\def\pion   {{\ensuremath{\Ppi}}\xspace}
\def\piz    {{\ensuremath{\pion^0}}\xspace}

\def\pip    {{\ensuremath{\pion^+}}\xspace}
\def\pim    {{\ensuremath{\pion^-}}\xspace}

\def\kaon    {{\ensuremath{\PK}}\xspace}
  \def\Kbar    {{\kern 0.2em\overline{\kern -0.2em \PK}{}}\xspace}

\def\KorKbar    {\kern 0.18em\optbar{\kern -0.18em K}{}\xspace}

\def\Kp      {{\ensuremath{\kaon^+}}\xspace}
\def\Km      {{\ensuremath{\kaon^-}}\xspace}

  \def\Dbar    {{\kern 0.2em\overline{\kern -0.2em \PD}{}}\xspace}
\def\D       {{\ensuremath{\PD}}\xspace}

\def\DorDbar    {\kern 0.18em\optbar{\kern -0.18em D}{}\xspace}
\def\Dz      {{\ensuremath{\D^0}}\xspace}

\def\Dp      {{\ensuremath{\D^+}}\xspace}

\def\Dstarp  {{\ensuremath{\D^{*+}}}\xspace}

\def\B       {{\ensuremath{\PB}}\xspace}
\def\Bbar    {{\ensuremath{\kern 0.18em\overline{\kern -0.18em \PB}{}}}\xspace}

\def\BorBbar    {\kern 0.18em\optbar{\kern -0.18em B}{}\xspace}
\def\Bz      {{\ensuremath{\B^0}}\xspace}

\def\Bu      {{\ensuremath{\B^+}}\xspace}
\def\Bub     {{\ensuremath{\B^-}}\xspace}
\def\Bp      {{\ensuremath{\Bu}}\xspace}
\def\Bm      {{\ensuremath{\Bub}}\xspace}

\def\Bs      {{\ensuremath{\B^0_\squark}}\xspace}

\def\jpsi     {{\ensuremath{{\PJ\mskip -3mu/\mskip -2mu\Ppsi\mskip 2mu}}}\xspace}

  \def\Y#1S{\ensuremath{\PUpsilon{(#1S)}}\xspace}

\def\Xires       {{\ensuremath{\PXi}}\xspace}

\def\Lz          {{\ensuremath{\PLambda}}\xspace}
\def\Lbar        {{\ensuremath{\kern 0.1em\overline{\kern -0.1em\PLambda}}}\xspace}
\def\LorLbar    {\kern 0.18em\optbar{\kern -0.18em \PLambda}{}\xspace}

\def\Sigmares    {{\ensuremath{\PSigma}}\xspace}

\def\Lb      {{\ensuremath{\Lz^0_\bquark}}\xspace}

\def\Lc      {{\ensuremath{\Lz^+_\cquark}}\xspace}

\def\Xibz    {{\ensuremath{\Xires^0_\bquark}}\xspace}
\def\Xibm    {{\ensuremath{\Xires^-_\bquark}}\xspace}

\def\Xicz    {{\ensuremath{\Xires^0_\cquark}}\xspace}
\def\Xicp    {{\ensuremath{\Xires^+_\cquark}}\xspace}

\def\to                 {\ensuremath{\rightarrow}\xspace}

\def\AT#1     {\ensuremath{A_{\mathrm{T}}^{#1}}\xspace}           

\def\C#1      {\ensuremath{\mathcal{C}_{#1}}\xspace}                       
\def\Cp#1     {\ensuremath{\mathcal{C}_{#1}^{'}}\xspace}                    
\def\Ceff#1   {\ensuremath{\mathcal{C}_{#1}^{\mathrm{(eff)}}}\xspace}        
\def\Cpeff#1  {\ensuremath{\mathcal{C}_{#1}^{'\mathrm{(eff)}}}\xspace}       
\def\Ope#1    {\ensuremath{\mathcal{O}_{#1}}\xspace}                       
\def\Opep#1   {\ensuremath{\mathcal{O}_{#1}^{'}}\xspace}                    

\newcommand{\tev}{\ifthenelse{\boolean{inbibliography}}{\ensuremath{~T\kern -0.05em eV}\xspace}{\ensuremath{\mathrm{\,Te\kern -0.1em V}}}\xspace}
\newcommand{\gev}{\ensuremath{\mathrm{\,Ge\kern -0.1em V}}\xspace}
\newcommand{\mev}{\ensuremath{\mathrm{\,Me\kern -0.1em V}}\xspace}
\newcommand{\kev}{\ensuremath{\mathrm{\,ke\kern -0.1em V}}\xspace}
\newcommand{\ev}{\ensuremath{\mathrm{\,e\kern -0.1em V}}\xspace}
\newcommand{\gevc}{\ensuremath{{\mathrm{\,Ge\kern -0.1em V\!/}c}}\xspace}
\newcommand{\mevc}{\ensuremath{{\mathrm{\,Me\kern -0.1em V\!/}c}}\xspace}
\newcommand{\gevcc}{\ensuremath{{\mathrm{\,Ge\kern -0.1em V\!/}c^2}}\xspace}
\newcommand{\gevgevcccc}{\ensuremath{{\mathrm{\,Ge\kern -0.1em V^2\!/}c^4}}\xspace}
\newcommand{\mevcc}{\ensuremath{{\mathrm{\,Me\kern -0.1em V\!/}c^2}}\xspace}

\def\mum  {\ensuremath{{\,\upmu\rm m}}\xspace}

\def\invfb   {\ensuremath{\mbox{\,fb}^{-1}}\xspace}

\def\gsim{{~\raise.15em\hbox{$>$}\kern-.85em
          \lower.35em\hbox{$\sim$}~}\xspace}
\def\lsim{{~\raise.15em\hbox{$<$}\kern-.85em
          \lower.35em\hbox{$\sim$}~}\xspace}

\def\pt         {\mbox{$p_{\rm T}$}\xspace}

\def\evtgen     {\mbox{\textsc{EvtGen}}\xspace}

\def\gauss      {\mbox{\textsc{Gauss}}\xspace}
\def\geant      {\mbox{\textsc{Geant4}}\xspace}

\def\photos     {\mbox{\textsc{Photos}}\xspace}

\def\pythia     {\mbox{\textsc{Pythia}}\xspace}

\def\tell1  {TELL1\xspace}
\def\ukl1   {UKL1\xspace}

\def\rstat{{\rm stat}}

\def\rsigma{r_{\sigma}}
\def\fsigma{f_{\sigma}}
\def\falphap{f_{\alpha_{+}}}
\def\falpham{f_{\alpha_{-}}}

\usepackage{cite} 
\usepackage{mciteplus}

\usepackage{longtable} 

\begin{document}

\renewcommand{\thefootnote}{\fnsymbol{footnote}}
\setcounter{footnote}{1}


\begin{titlepage}
\pagenumbering{roman}

\vspace*{-1.5cm}
\centerline{\large EUROPEAN ORGANIZATION FOR NUCLEAR RESEARCH (CERN)}
\vspace*{1.5cm}
\hspace*{-0.5cm}
\begin{tabular*}{\linewidth}{lc@{\extracolsep{\fill}}r}
\ifthenelse{\boolean{pdflatex}}
{\vspace*{-2.7cm}\mbox{\!\!\!\includegraphics[width=.14\textwidth]{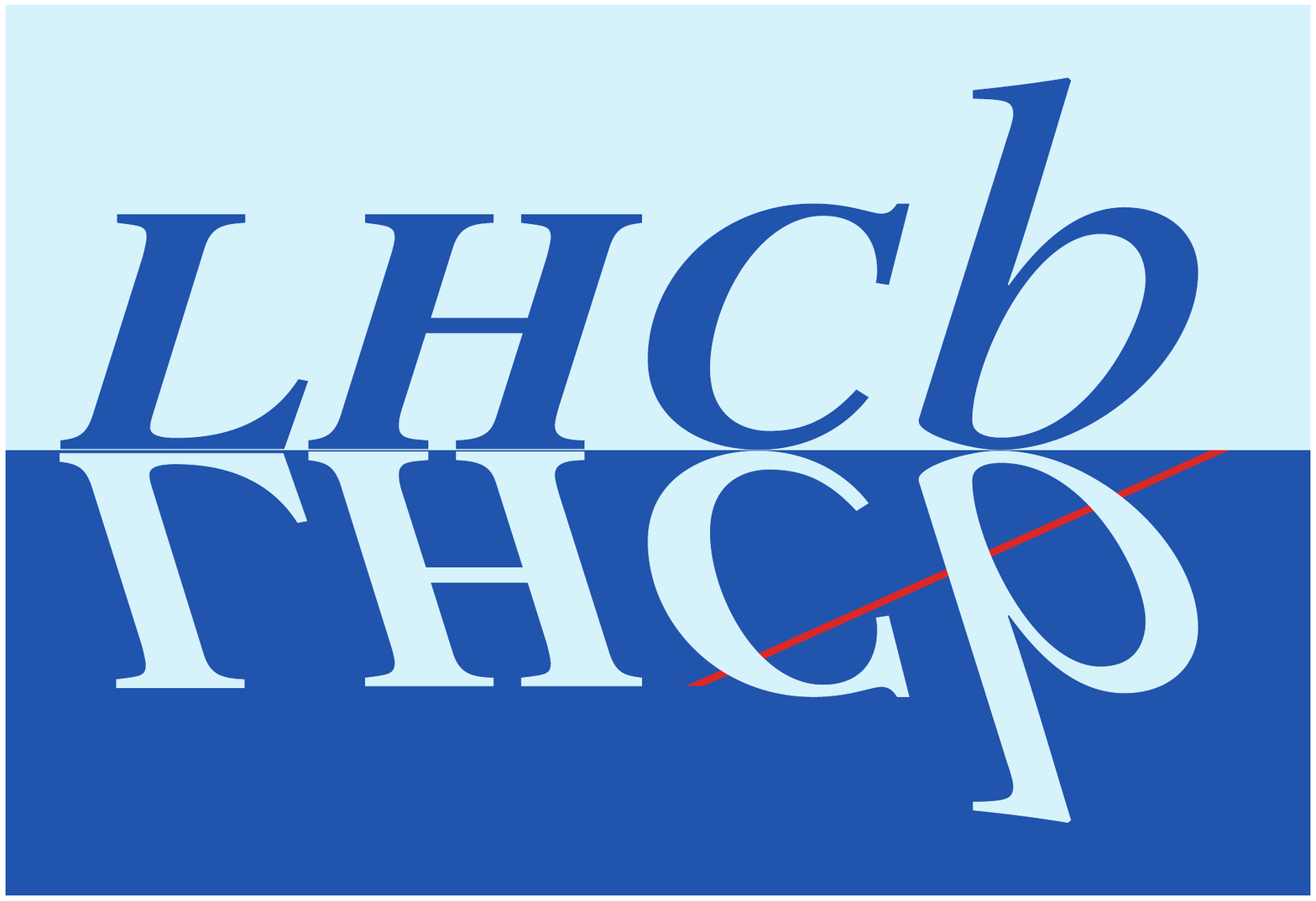}} & &}%
{\vspace*{-1.2cm}\mbox{\!\!\!\includegraphics[width=.12\textwidth]{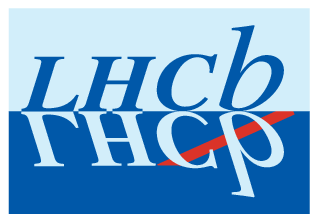}} & &}%
\\
 & & CERN-PH-EP-2014-226 \\  
 & & LHCb-PAPER-2014-048 \\  
 & & Sept. 30, 2014 \\ 
 & & \\
\end{tabular*}

\vspace*{1.0cm}

{\bf\boldmath\huge
\begin{center}
  Precision measurement of the mass and lifetime of the $\Xibm$ baryon
\end{center}
}

\vspace*{2.0cm}

\begin{center}
The LHCb collaboration\footnote{Authors are listed at the end of this Letter.}
\end{center}

\vspace{\fill}

\begin{abstract}
  \noindent
We report on measurements of the mass and lifetime of the $\Xibm$ baryon
using about 1800 $\Xibm$ decays reconstructed in a proton-proton collision data set 
corresponding to an integrated luminosity of 3.0\invfb collected by the LHCb experiment.
The decays are reconstructed in the $\Xibm\to\Xicz\pim$, $\Xicz\to p\Km\Km\pip$ channel
and the mass and lifetime are measured using the $\Lb\to\Lc\pim$ mode as a reference. We measure 
\begin{align*}
M(\Xibm) - M(\Lb) &= 178.36\pm0.46\pm0.16~\mevcc, \\ 
\frac{\tau_{\Xibm}}{\tau_{\Lb}} &= 1.089\pm0.026\pm0.011, 
\end{align*}
\noindent where the uncertainties are statistical and systematic, respectively. 
These results lead to a factor of two better precision 
on the $\Xibm$ mass and lifetime compared to previous best measurements, and are consistent with 
theoretical expectations.
\end{abstract}

\vspace*{1.0cm}

\begin{center}
  Published in Physical Review Letters
\end{center}

\vspace{\fill}

{\footnotesize 
\centerline{\copyright~CERN on behalf of the \lhcb collaboration, license \href{http://creativecommons.org/licenses/by/4.0/}{CC-BY-4.0}.}}
\vspace*{2mm}

\end{titlepage}


\newpage
\setcounter{page}{2}
\mbox{~}


\cleardoublepage


\renewcommand{\thefootnote}{\arabic{footnote}}
\setcounter{footnote}{0}



\pagestyle{plain} 
\setcounter{page}{1}
\pagenumbering{arabic}


%


Over the last two decades, beauty mesons have been studied in detail. Various theoretical approaches
allow one to relate measured decay rates to Standard Model parameters. 
One of the most predictive tools is the heavy quark expansion
(HQE)~\cite{Khoze:1983yp,Bigi:1991ir,Bigi:1992su,Blok:1992hw,Blok:1992he,Neubert:1997gu,Uraltsev:1998bk,Bigi:1995jr}, which
describes the decay rates of beauty hadrons through an expansion in powers of $\Lambda_{\rm QCD}/m_b$, where $\Lambda_{\rm QCD}$ 
is the energy scale at which the strong-interaction coupling becomes large, and $m_b$ is the $b$-quark mass.
In addition to the total $b$-hadron decay widths, HQE can be used to calculate $b$-hadron parameters required for  
the measurement of coupling strengths between quarks in charged-current interactions, which in
turn provides constraints on physics beyond the Standard Model.

A stringent test of HQE is to confront its predictions for lifetimes, {\it i.e.,} the inverse of the corresponding 
decay widths, with precision measurements. The lifetimes of the $\Bz$ and $\Bp$ mesons are measured 
to a precision of about 0.5\%~\cite{PDG2014}, the $\Bs$ meson to 1\%~\cite{LHCb-PAPER-2014-037,PDG2014},
and the $\Lb$ baryon to 0.7\%~\cite{PDG2014}, and their values are in 
agreement with HQE predictions~\cite{Stone:Lifetimes}.

Another interesting test is to compare the measured lifetime ratio $\tau(\Xibm)/\tau(\Xibz)$ to HQE
predictions. Since penguin contraction terms cancel in this ratio~\cite{Lenz:LHQE2014}, a more precise
prediction is possible compared to $\tau(\Lb)/\tau(\Bz)$.
One prediction leads to $\tau(\Xibm)/\tau(\Xibz)=1.05\pm0.07$~\cite{Lenz:LHQE2014}, 
where the dominant uncertainties are
related to matrix elements that are calculable using lattice quantum chromodynamics (QCD)~\cite{Wilson:1974sk}. 
A phenomenological analysis of the relevant matrix elements using charm baryon lifetimes leads to a prediction of
$1/\tau(\Lb)-1/\tau(\Xibm)=0.11\pm0.03$\,ps$^{-1}$~\cite{Voloshin:1999ax}, or $\tau(\Xibm)/\tau(\Lb) = 1.19^{+0.07}_{-0.06}$.
Recently, the first measurement of the lifetime ratio $\tau(\Xibz)/\tau(\Lb)$ was made,
yielding $\tau(\Xibz)/\tau(\Lb) = 1.006\pm0.018\pm0.010$~\cite{LHCb-PAPER-2014-021}. Previous $\Xibm$ lifetime measurements,
which used $\Xibm\to\jpsi\Xires^-$ decays, led to values of $1.55^{+0.10}_{-0.09}\pm0.03$~ps~\cite{LHCb-PAPER-2014-010} and
$1.32\pm0.14\pm0.02$~ps~\cite{Aaltonen:2014wfa}. The weighted average of these 
two results, along with the recent $\Xibz$ lifetime measurement~\cite{LHCb-PAPER-2014-021},
yields $\tau(\Xibm)/\tau(\Xibz)=1.00\pm0.06$. Improved experimental and theoretical precision 
of the $\Xibm$ lifetime will allow for a more stringent test of the HQE prediction.

Measurements of $b$-baryon masses and isospin splittings provide information on the interquark potential.
A number of QCD-inspired models predict  the $\Xibz$ and $\Xibm$ masses, or their average, 
which range from approximately
5780\mevcc to 5900\mevcc~\cite{oai:arXiv.org:hep-ph/0504112,oai:arXiv.org:0712.0406,oai:arXiv.org:0710.0123,oai:arXiv.org:hep-ph/9502251,Karliner:2008sv,oai:arXiv.org:0806.4951,Roberts:2007ni,Ghalenovi,Patel:2007gx,Patel:2008nv}. 
More accurate predictions exist for the $\Xibm-\Xibz$ mass splitting, estimated 
to be $6.24\pm0.21$\mevcc or $6.4\pm1.6$\mevcc when
extrapolating from the measured isospin splitting $M(\Xires^-)-M(\Xires^0)$ or $M(\Xicz)-M(\Xicp)$, 
respectively~\cite{Karliner:2008sv}.
The $\Xibm$ mass is currently known to a precision of 1.0\mevcc~\cite{LHCb-PAPER-2012-048}, which is a factor of 
three less precise than that of the $\Xibz$ baryon~\cite{LHCb-PAPER-2014-021}.

In this Letter, we report improved measurements of the mass and lifetime of the $\Xibm$ baryon using
about 1800 $\Xibm\to\Xicz\pim$, $\Xicz\to p\Km\Km\pip$ signal decays. 
The measurements are normalized using the $\Lb\to\Lc\pim$, $\Lc\to p\Km\pip$ decay as a reference. 
Charge conjugate processes are implied throughout. 

The measurements use proton-proton ($pp$) collision data samples, collected by the LHCb experiment,
corresponding to an integrated luminosity of 3.0\invfb, of which 1.0\invfb was recorded at 
a center-of-mass energy of 7\tev and 2.0\invfb at 8\tev.  
The \lhcb detector~\cite{Alves:2008zz} is a single-arm forward
spectrometer covering the \mbox{pseudorapidity} range $2<\eta <5$,
designed for the study of particles containing \bquark or \cquark
quarks. The detector includes a high-precision tracking system,
which provides a momentum measurement with precision of about 0.5\% from
2$-$100~\gevc and impact parameter resolution of 20\mum for
particles with large transverse momentum (\pt). The polarity of the dipole magnet is 
reversed periodically throughout data-taking to reduce asymmetries in the detection of charged particles.
Ring-imaging Cherenkov detectors~\cite{LHCb-DP-2012-003}
are used to distinguish charged hadrons. Photon, electron and
hadron candidates are identified using a calorimeter system, followed by
detectors to identify muons~\cite{LHCb-DP-2012-002}.

The trigger~\cite{LHCb-DP-2012-004} consists of a
hardware stage, based on information from the calorimeter and muon
systems, followed by a software stage, which applies a full event
reconstruction~\cite{LHCb-DP-2012-004,BBDT}. 
About 57\% of the selected $X_b$ events are triggered at the hardware 
level by one or more of the $X_b$ final-state particles.
(Throughout, we use $X_b$ ($X_c$) to refer to either a $\Xibm$ ($\Xicz$) or $\Lb$ ($\Lc$) baryon.)
The remaining 43\% are triggered only on other activity in the event.
We refer to these two classes of events as triggered on signal (TOS) and
triggered independently of signal (TIS).
The software trigger requires a two-, three- or four-track
secondary vertex with a large scalar \pt sum of
the particles and a significant displacement from the primary $pp$
interaction vertices~(PVs). At least one particle should have $\pt>1.7\gevc$ and 
be inconsistent with coming from any of the PVs. The signal candidates 
are required to pass a multivariate software trigger
selection algorithm~\cite{BBDT}. 

Proton-proton collisions are simulated using
\pythia~\cite{Sjostrand:2006za,*Sjostrand:2007gs} with a specific \lhcb
configuration~\cite{LHCb-PROC-2010-056}.  Decays of hadronic particles
are described by \evtgen~\cite{Lange:2001uf}, in which final-state
radiation is generated using \photos~\cite{Golonka:2005pn}. The
interaction of the generated particles with the detector and its
response are implemented using the \geant toolkit~\cite{Allison:2006ve, *Agostinelli:2002hh} as described in
Ref.~\cite{LHCb-PROC-2011-006}. The $X_c$ final states are modeled using a combination of
resonant and nonresonant contributions to reproduce the substructures seen in data.

Signal $\Xibm$ ($\Lb$) candidates are formed by combining in a kinematic fit a 
$\Xicz\to p\Km\Km\pip$ ($\Lc\to p\Km\pip$) candidate with
a $\pim$ candidate (referred to as the bachelor). 
The $X_b$ candidate is included in the fit to each PV and is then associated with the one for which the $\chi^2$ increases by the 
smallest amount. The kinematic fit exploits PV, $X_b$ and $X_c$ decay-vertex constraints to improve the mass resolution. 
The $X_c$ decay products are each required to have $\pt>100$\mevc, and the bachelor pion is required to have $\pt>500$\mevc.
All final-state particles from the signal candidate are required to have trajectories that are 
significantly displaced from the PV and to pass particle identification (PID) requirements.
The $\Km$ and $\pip$ PID efficiencies are determined from $\Dstarp\to\Dz\pip$, $\Dz\to\Km\pip$
calibration samples, whereas the proton PID efficiency is determined from simulation.
The PID efficiencies are reweighted to account for different momentum spectra and track occupancies between
the calibration and signal samples. The efficiencies of the PID requirements on the $\Xicz$ and $\Lc$ final states
are 80\% and 86\%, respectively. Mass vetoes are used to suppress cross-feeds from misidentified $D^{+}_{(s)}\to\Kp\Km\pip$,
$D^{*+}\to\Dz(\Kp\Km)\pip$, and $\Dp\to\Km\pip\pip$ decays faking $\Lc\to p\Km\pip$ decays, as in Ref.~\cite{LHCb-PAPER-2014-021}.
The difference between the $\Xicz$ ($\Lc$) candidate mass and the known value~\cite{PDG2014}
is required to be less than 14\mevcc (20\mevcc), which is about 2.5 times
the mass resolution.

To improve the signal-to-background ratio, we employ a boosted decision tree (BDT) discriminant~\cite{Breiman, AdaBoost} 
built from the same variables
used in Ref.~\cite{LHCb-PAPER-2014-021}. 
To train the BDT, the kinematic distributions of the 
signal are modeled using simulated decays. The background is modeled using signal candidates with
$X_b$ invariant mass greater than 300\mevcc above the signal peak mass. To increase the size of the background sample for the
$\Xibm$ BDT training, we also include events in the $\Xicz$ sideband regions, $20 < |M(p\Km\Km\pip)-M(\Xicz)| < 50$\mevcc.
The BDT requirement is chosen to minimize the expected $\Xibm$ relative yield uncertainty, corresponding
to a selection efficiency of 97\% (50\%) for signal (combinatorial background).
The fraction of events with multiple candidates is below 1\% (mostly one extra candidate) over the full fit range
in both the signal and normalization modes. All candidates are kept.

The invariant mass signal shapes are obtained from simulated $\Xibm\to\Xicz\pim$ and $\Lb\to\Lc\pim$ decays. They
are each modeled by the sum of two Crystal Ball (CB) functions~\cite{Skwarnicki:1986xj}
with a common mean as
\begin{align}
f_{\rm sig}^{\Lb} &= f_{\rm low}\times {\rm CB}_{-}(m_0,\sigma_{-},\alpha_{-},n) + (1-f_{\rm low})\times{\rm CB}_{+}(m_0,\sigma_{+},\alpha_{+},n) \\ 
f_{\rm sig}^{\Xibm} &= f_{\rm low}\times {\rm CB}_{-}(m_0^{\prime},\fsigma\sigma_{-},\falpham\alpha_{-},n) + (1-f_{\rm low})\times{\rm CB}_{+}(m_0^{\prime},\fsigma\sigma_{+},\falphap\alpha_{+},n). 
\end{align}
\noindent The CB functions each include a Gaussian component to describe the core of the mass distribution,
as well as power-law tails to describe the radiative tail below (CB$_-$) and the non-Gaussian resolution above (CB$_+$) 
the signal peak. The extent of these 
tails is governed by the width and tail parameters, $\sigma_{\pm}$ and $\alpha_{\pm}$, respectively.
The parameter $m_0$ is the fitted $\Lb$ mass, and $m_0^{\prime}\equiv m_0+\delta M$ is the $\Xibm$ mass, written in terms of
the fitted mass difference, $\delta M$, between the two signals. The low-mass CB width, $\sigma_{-}$, is expressed in terms 
of the high-mass width using $\sigma_{-}=\rsigma\sigma_{+}$.
The parameters $\fsigma$ and $f_{\alpha_{\pm}}$ allow for possible differences in the mass resolutions and tail parameters, respectively,
between the signal and normalization modes. 
We fix the power $n=10$ and $f_{\rm low}=0.5$ to minimize the number of correlated parameters in the signal shape.
The parameters $\rsigma$, $\falphap$, $\falpham$, and $\fsigma$ are determined from simulated decays, and they
are consistent with unity. These four parameters are fixed in fits to the data to the values from simulation, while
$\sigma_{+}$, $\alpha_{+}$ and $\alpha_{-}$ are freely varied, along with $m_0$ and $\delta M$.

The invariant mass spectra also include partially reconstructed $b$-baryon background 
contributions, misidentified $\Km$ in $X_b\to X_c\Km$ decays, as well as random track combinations, 
primarily from false $X_c$ candidates. The main source of partially reconstructed background is from
$X_b\to X_c\rho^-$ decays, where a $\piz$  from the $\rho^-$ decay is not used to form the candidate. Its shape is
obtained from simulated $\Lb\to\Lc\rho^-$ decays, and is assumed to be the same for both
the signal and normalization modes, apart from a shift in the overall mass spectrum. 
A contribution from $\Lb\to\Sigmares_c^+\pim,~\Sigmares_c^+\to\Lc\piz$ decays 
is also expected to populate the $\Lc\pim$ mass spectrum, and its shape is taken to be the same
to that of the $\Lb\to\Lc\rho^-$ signal. An additional contribution from partially reconstructed $\Xires_b$ decays
is found, through a study of the $\Lc$ sidebands, to populate the $\Lc\pim$ mass spectrum. 
This background is modeled through a fit to the
$\Lb$ candidate mass spectrum obtained using the lower and upper $\Lc$ mass sidebands. 
The shape of the background from misidentified $X_b\to X_c\Km$ decays
is taken from simulation. The misidentification rate of 3.1\% is obtained from $\Dstarp\to\Dz\pip$
calibration samples, reweighted in \pt, $\eta$ and number of tracks to match the distributions observed in data.
No peaking contributions from charmless backgrounds are observed when studying the $X_b$ mass spectra using 
the $X_c$ mass sidebands. The combinatorial background is modeled using an exponential function with a freely varying
slope.

\begin{figure}[tb]
\centering
\includegraphics[width=0.49\textwidth]{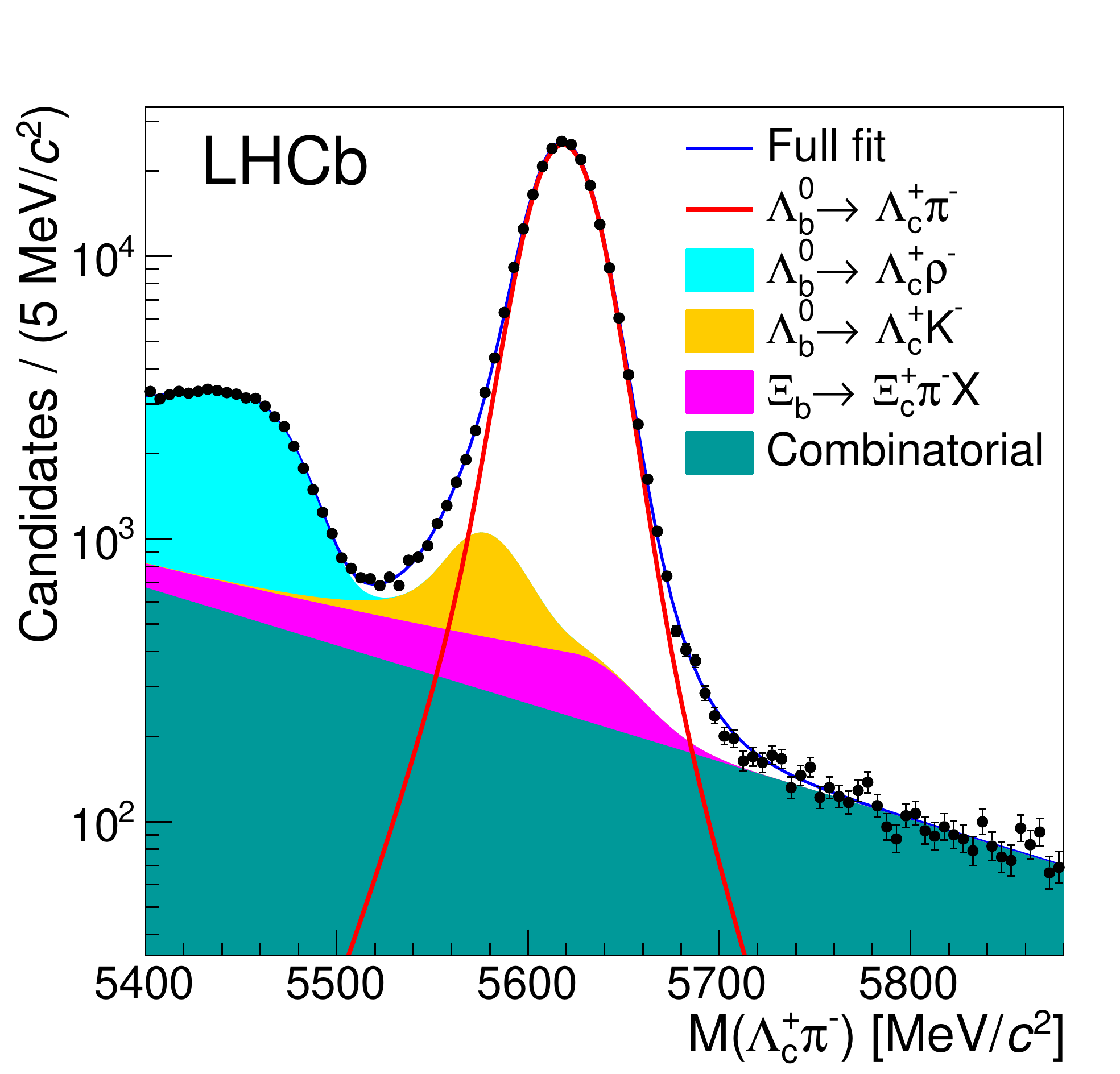}
\includegraphics[width=0.49\textwidth]{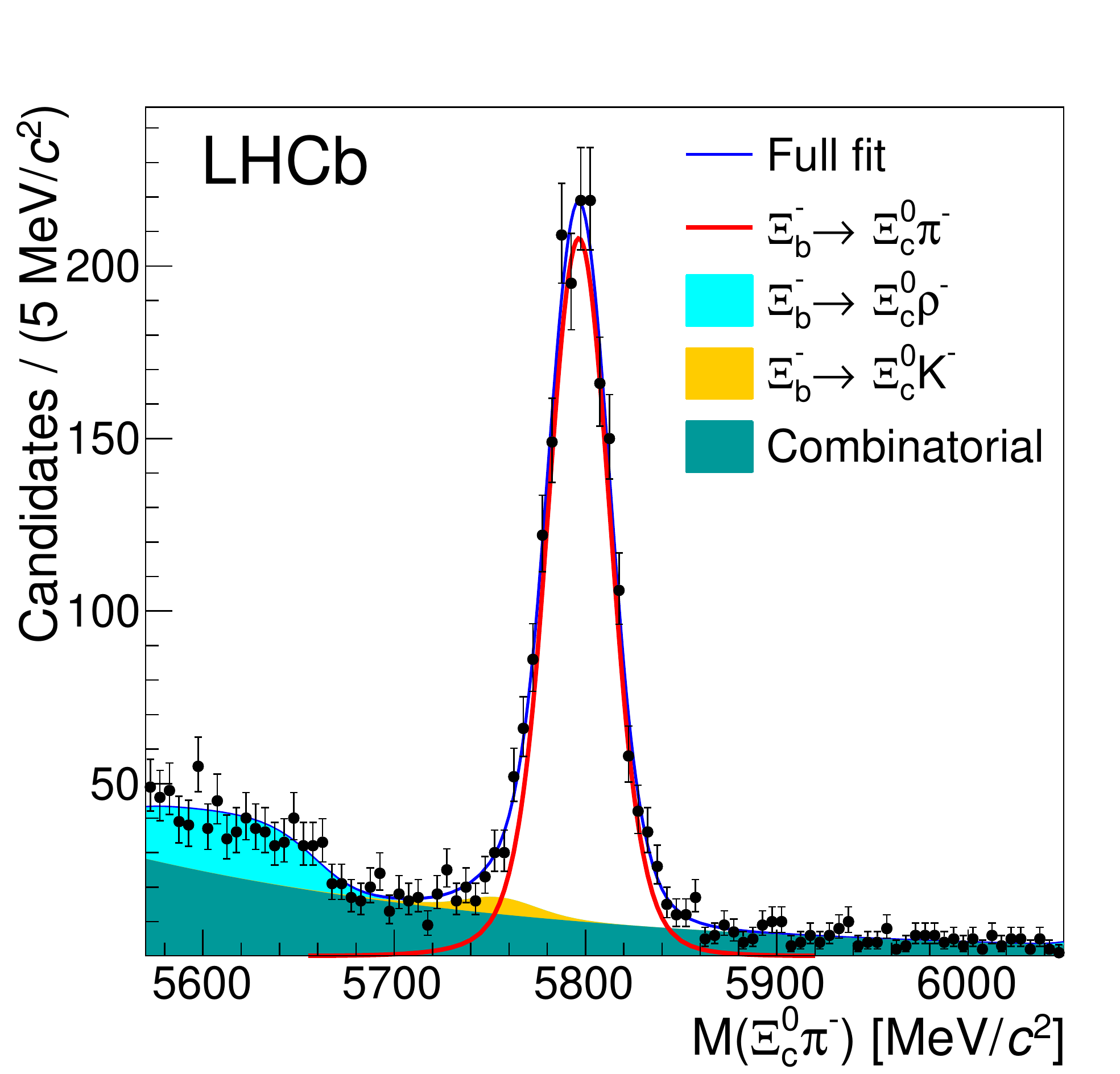}
\caption{\small{Invariant mass spectrum, along with the fit projections, for (left) $\Lb\to\Lc\pim$ and
(right) $\Xibm\to\Xicz\pim$ candidates.}}
\label{fig:LbNormFits}
\end{figure}

The $\Lc\pim$ and $\Xicz\pim$ mass spectra are fit simultaneously using a binned maximum likelihood fit. 
The results of the fit are shown in Fig.~\ref{fig:LbNormFits}. A total of $1799\pm46$ $\Xibm\to\Xicz\pim$ and
$(220.0\pm0.5)\times10^{3}$ $\Lb\to\Lc\pim$ signal decays are observed. The mass difference is
measured to be 
\begin{align*}
\delta M\equiv M(\Xibm) - M(\Lb) &= 178.36\pm0.46\mevcc,
\end{align*}
\noindent where the uncertainty is statistical only. 

The observed signals are also used to measure the $\Xibm$ baryon lifetime relative to that of the $\Lb$ baryon.
We measure the efficiency-corrected yields in six bins of measured decay time, as given in Table~\ref{tab:timeFits_Lb}.
The ratio of efficiency-corrected yields depends exponentially on decay time
as $N_{\rm cor}[{\Xibm\to\Xicz\pim}](t)/N_{\rm cor}[{\Lb\to\Lc\pim}](t) = e^{\beta t}$, where 
$\beta = 1/\tau(\Lb)-1/\tau(\Xibm)$. Many systematic uncertainties cancel
to first order in the ratio, such as those associated with the time resolutions and relative acceptances. 

\begin{figure}[tb]
\centering
\includegraphics[width=0.8\textwidth]{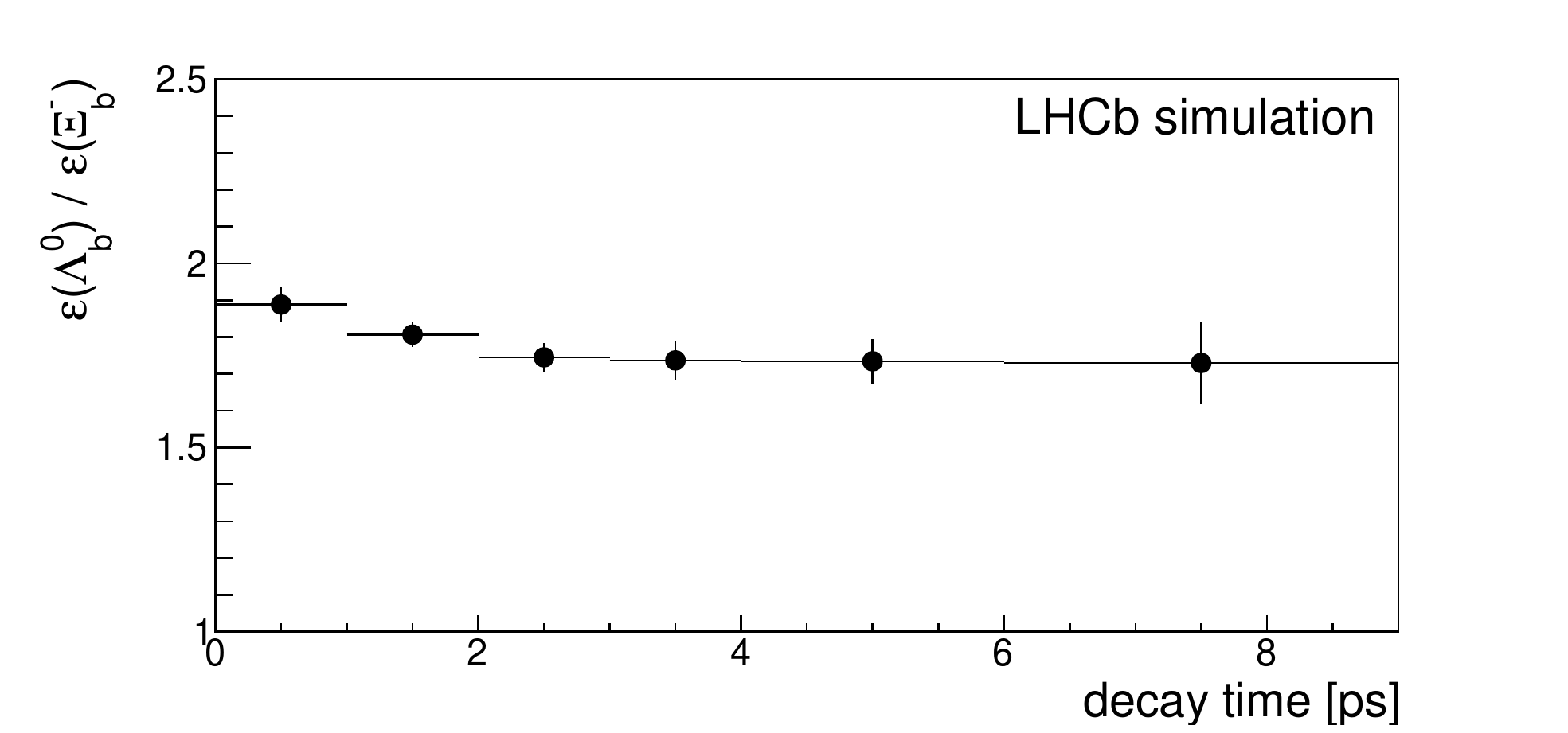}
\caption{\small{Ratio of the $\Lb\to\Lc\pim$ to the $\Xibm\to\Xicz\pim$ selection efficiencies
as a function of decay time. The uncertainties are due to the finite size of the simulated samples.}}
\label{fig:TauEff}
\end{figure}

The yields in each time bin are obtained using the results from the full fit with the signal shape parameters fixed.
No dependence of the signal shapes on decay time is observed in simulated decays, as expected.
The background shape parameters are also fixed, except for the combinatorial background shape parameter, and one of
the $X_b\to X_c\rho$ shape parameters, which is seen to have a dependence on decay time.
The signal yields in each of the time bins are shown in Table~\ref{tab:timeFits_Lb}.
\begin{table*}[tb]
\begin{center}
\caption{\small{Fitted yields of $\Lb\to\Lc\pim$ and $\Xibm\to\Xicz\pim$ in each time bin. Uncertainties are statistical only.}}
\begin{tabular}{lcc}
\hline\hline
Decay time (ps)            &   $\Lb\to\Lc\pim$    & $\Xibm\to\Xicz\pim$ \\
\hline
$0-1$                  &   $38,989\pm212$ & $260\pm17$ \\
$1-2$                  &   $79,402\pm299$ & $629\pm27$ \\
$2-3$                  &   $48,979\pm233$ & $436\pm22$ \\
$3-4$                  &   $26,010\pm169$ & $232\pm16$ \\
$4-6$                  &   $19,651\pm147$ & $177\pm14$ \\
$6-9$                  &   $5794\pm79$    & $69\pm9$ \\
\hline\hline
\end{tabular}
\label{tab:timeFits_Lb}
\end{center}
\end{table*}
The relative acceptance, shown in Fig.~\ref{fig:TauEff}, is obtained using simulated decays after applying all event
selection criteria.
The efficiency for reconstructing the $\Xibm\to\Xicz\pim$ mode is about a factor of two lower than that of 
the $\Lb\to\Lc\pim$ decay due to the extra particle in the final state and the 
lower average momentum of the final-state particles.
The relative efficiency, $\epsilon(\Lb)/\epsilon(\Xibm)$, is nearly uniform, with a gradual increase 
for decay times below 2~ps. 
This increase is expected, because the $\Lc$ lifetime is about twice that of the 
$\Xicz$ baryon, and the correspondingly larger impact parameters are favored by the software trigger 
and offline selections, most notably when the $X_b$ decay time is small.

\begin{figure}[tb]
\centering
\includegraphics[width=0.98\textwidth]{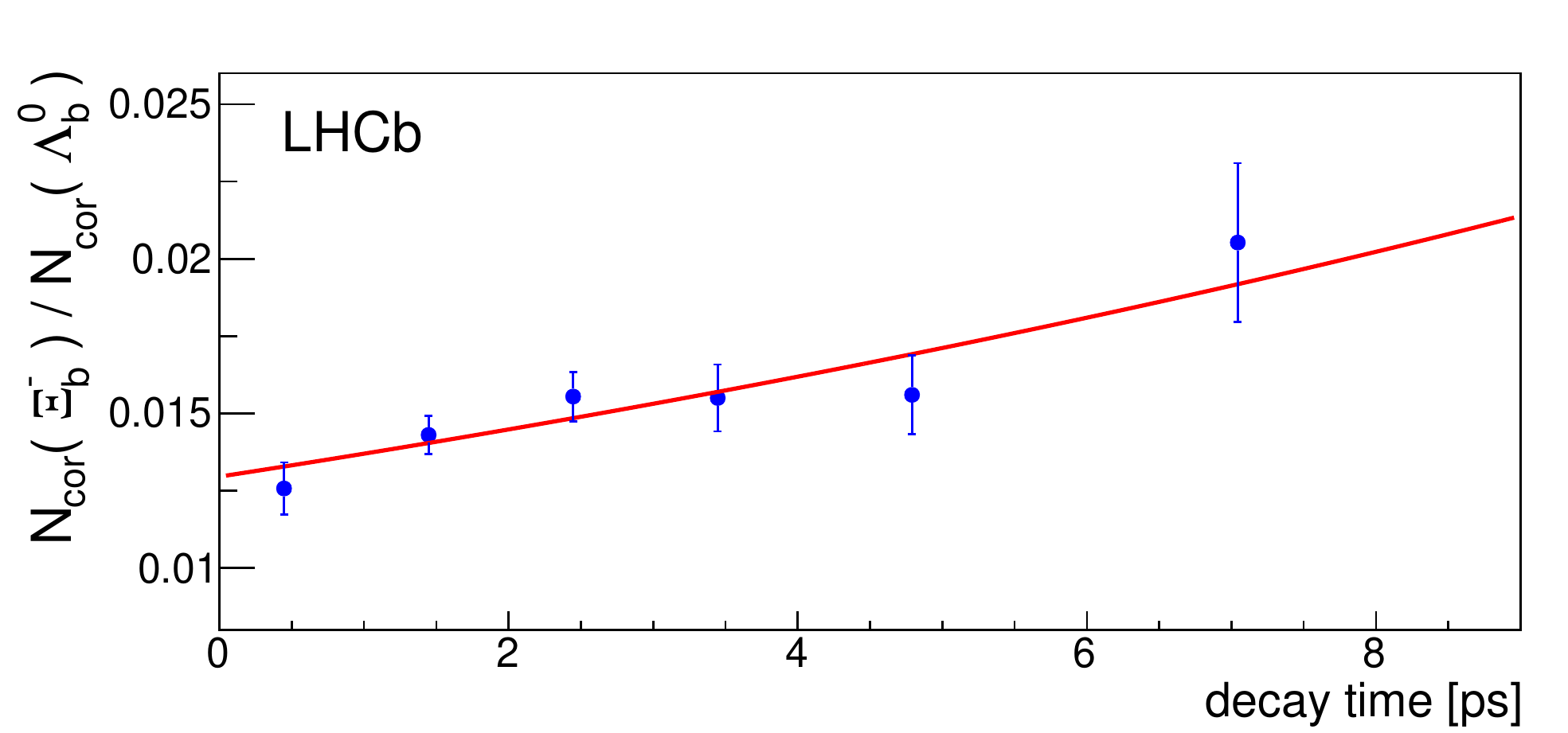}
\caption{\small{Corrected yield ratio, $N_{\rm cor}(\Xibm)/N_{\rm cor}(\Lb)$ 
in bins of decay time, along with the exponential fit. The uncertainties are statistical only.}}
\label{fig:CorrYieldRatio}
\end{figure}

The ratios of corrected yields and the exponential fit are shown in Fig.~\ref{fig:CorrYieldRatio}. 
The points are displayed at the average time value in the bin assuming
an exponential time distribution with mean 1.54 ps, which is the average of the known
$\Lb$ and fitted $\Xibm$ lifetimes. 
Choosing either the $\Lb$ or the fitted $\Xibm$ lifetime leads to a negligible change in the result.
The fitted value is $\beta = 0.0557\pm0.0160$~ps$^{-1}$, where
the uncertainty is statistical only. 
Using $\tau(\Lb)=1.468\pm0.009\pm0.008$~ps~\cite{LHCb-PAPER-2014-003}, we find
\begin{align*}
r_{\tau}\equiv\frac{\tau_{\Xibm}}{\tau_{\Lb}} &= 1.089\pm0.026\,(\rstat). 
\end{align*}

Several consistency checks are performed,
including comparing the mass differences obtained from 7\tev versus 8\tev data, 
opposite magnet polarities, $X_b$ versus $\overline{X}_b$ samples, and different trigger selections.
In all cases, the results are consistent with statistical fluctuations of independent samples.
In addition, the analysis is carried out using
15,500 $\Bm\to\Dz\pim$, $\Dz\to\Km\Kp\pip\pim$ signal decays for normalization.
The $\Xibm$ mass and lifetime results agree with the above values to better than one 
standard deviation, considering only the uncertainty due to the $\Lb$ and $\Bm$ masses and lifetimes.

The measurements of $M(\Xibm)$ and $\tau(\Xibm)$ are subject to systematic uncertainties,
but the largest contributions cancel to first order in $\delta M$ and $r_{\tau}$. 
For the mass difference measurement,
the effect of the momentum scale uncertainty of 0.03\%~\cite{LHCb-PAPER-2013-011} is investigated by shifting
the momenta of all final-state particles in simulated decays by this amount, leading to an uncertainty on 
$\delta M$ of 0.08\mevcc. Because the signal mode has one 
more particle than the normalization mode, the correction for energy loss in the detector material leads to
an additional uncertainty of 0.06\mevcc~\cite{LHCb-PAPER-2013-011}. Uncertainty due to the
signal modeling is 0.06\mevcc, obtained by shifting all fixed parameters by their uncertainties, and
adding the shifts in $\delta M$ from the nominal value in quadrature.
For the background model, several variations from
the nominal fit are investigated, including (a) using a second-order polynomial to describe the
combinatorial background, (b) allowing the fixed parameters in the partially reconstructed background to 
vary, (c) removing the $\Xires_{b}$ background component, (d) a 20\% relative increase in
the $\Xibm\to\Xicz\Km$ cross-feed, and (e) varying the fit range.
The changes in $\delta M$ are added in quadrature to obtain the background uncertainty of 0.11\mevcc.
Adding all sources of uncertainty in quadrature leads to a systematic uncertainty in $\delta M$ of 0.16\mevcc.

The largest source of systematic uncertainty in $r_{\tau}$ is the
limited size of the simulated samples, which contributes an uncertainty of 0.010.
The simulated efficiencies are averaged over TOS and TIS events in the simulation, 
of which the former comprises 67\% of the sample, compared to 57\% in data.
While the values of $r_{\tau}$ are statistically compatible between these two samples, if 
the efficiencies from simulation are reweighted to match the composition observed in data,
a change in $r_{\tau}$ of 0.004 is found. This shift is assigned as a systematic uncertainty.
Variation in the signal and background models lead to a negligible change in $r_{\tau}$.
We also consider possible different performances of the BDT in data versus simulation by
correcting the data with an efficiency obtained with a tighter BDT requirement. The difference
of 0.001 is assigned as a systematic uncertainty. For the proton efficiency, we use the values 
obtained from simulation. By varying the proton PID requirements, a maximal change of 0.001
is found, which is assigned as a 
systematic uncertainty. To investigate possible effects due to the larger $\Lc$ lifetime 
(than the $\Xicz$), we reject candidates with $ct$ larger than 150~$\mu$m. The difference of 0.003
from the nominal result is assigned as a systematic uncertainty. In total, the systematic
uncertainty on $r_{\tau}$ is 0.011.

In summary, we use a $pp$ collision data sample corresponding to 3.0\invfb of integrated luminosity
to improve the precision
of the $\Xibm$ mass and lifetime by a factor of two over the previous best measurements. 
The resulting mass difference and relative lifetime are
\begin{align*}
M(\Xibm)-M(\Lb) &= 178.36\pm0.46\pm0.16\mevcc, \\
\frac{\tau_{\Xibm}}{\tau_{\Lb}} &= 1.089\pm0.026\pm0.011, 
\end{align*}
\noindent where the uncertainties are statistical and systematic, respectively.
Using the measured $\Lb$ mass~\cite{LHCb-PAPER-2014-002} and lifetime~\cite{LHCb-PAPER-2014-003}, 
we find
\begin{align*}
M(\Xibm) &= 5797.72\pm0.46\pm0.16\pm0.26_{\Lb}~\mevcc,\\
\tau_{\Xibm} &= 1.599\pm0.041\pm0.018\pm0.012_{\Lb}~{\rm ps},
\end{align*}
\noindent where the last uncertainty is due to the precision on the $\Lb$ lifetime.
Using the measurements of the $\Xibz$ mass difference and relative lifetime,
$M(\Xibz)-M(\Lb) = 172.44\pm0.39\,\pm0.17$\mevcc and 
$\tau_{\Xibz}/\tau_{\Lb} = 1.006\pm0.018\pm0.010$~\cite{LHCb-PAPER-2014-021}, we obtain
\begin{align*}
M(\Xibm)-M(\Xibz) &= 5.92\pm0.60\pm0.23\mevcc \\
\frac{\tau_{\Xibm}}{\tau_{\Xibz}} &= 1.083\pm0.032\pm0.016. 
\end{align*} 
\noindent The measured isospin splitting between the $\Xibm$ and $\Xibz$ baryons is consistent with the prediction 
in Ref.~\cite{Karliner:2008sv} of $6.24\pm0.21$\mevcc. The relative lifetime is 2.3 standard deviations
larger than unity, giving a first indication that the $\Xibm$ baryon lifetime is larger than that of the $\Xibz$ baryon.
This result is consistent with 
the theoretical expectations of $\tau_{\Xibm}/\tau_{\Xibz}=1.05\pm0.07$~\cite{Lenz:LHQE2014} 
and $\tau_{\Xibm}/\tau_{\Lb}=1.19^{+0.07}_{-0.06}$~\cite{Voloshin:1999ax}, based on the HQE.


\section*{Acknowledgements}
\noindent We express our gratitude to our colleagues in the CERN
accelerator departments for the excellent performance of the LHC. We
thank the technical and administrative staff at the LHCb
institutes. We acknowledge support from CERN and from the national
agencies: CAPES, CNPq, FAPERJ and FINEP (Brazil); NSFC (China);
CNRS/IN2P3 (France); BMBF, DFG, HGF and MPG (Germany); SFI (Ireland); INFN (Italy); 
FOM and NWO (The Netherlands); MNiSW and NCN (Poland); MEN/IFA (Romania); 
MinES and FANO (Russia); MinECo (Spain); SNSF and SER (Switzerland); 
NASU (Ukraine); STFC (United Kingdom); NSF (USA).
The Tier1 computing centres are supported by IN2P3 (France), KIT and BMBF 
(Germany), INFN (Italy), NWO and SURF (The Netherlands), PIC (Spain), GridPP 
(United Kingdom).
We are indebted to the communities behind the multiple open 
source software packages on which we depend. We are also thankful for the 
computing resources and the access to software R\&D tools provided by Yandex LLC (Russia).
Individual groups or members have received support from 
EPLANET, Marie Sk\l{}odowska-Curie Actions and ERC (European Union), 
Conseil g\'{e}n\'{e}ral de Haute-Savoie, Labex ENIGMASS and OCEVU, 
R\'{e}gion Auvergne (France), RFBR (Russia), XuntaGal and GENCAT (Spain), Royal Society and Royal
Commission for the Exhibition of 1851 (United Kingdom).




\ifx\mcitethebibliography\mciteundefinedmacro
\PackageError{LHCb.bst}{mciteplus.sty has not been loaded}
{This bibstyle requires the use of the mciteplus package.}\fi
\providecommand{\href}[2]{#2}

\newpage

\centerline{\large\bf LHCb collaboration}
\begin{flushleft}
\small
R.~Aaij$^{41}$, 
B.~Adeva$^{37}$, 
M.~Adinolfi$^{46}$, 
A.~Affolder$^{52}$, 
Z.~Ajaltouni$^{5}$, 
S.~Akar$^{6}$, 
J.~Albrecht$^{9}$, 
F.~Alessio$^{38}$, 
M.~Alexander$^{51}$, 
S.~Ali$^{41}$
% ...

{\footnotesize \it
% ...
$ ^{5}$Clermont Universit\'{e}, Universit\'{e} Blaise Pascal, CNRS/IN2P3, LPC, Clermont-Ferrand, France\\
$ ^{6}$CPPM, Aix-Marseille Universit\'{e}, CNRS/IN2P3, Marseille, France\\
% ...
$ ^{9}$Fakult\"{a}t Physik, Technische Universit\"{a}t Dortmund, Dortmund, Germany\\
% ...
$ ^{37}$Universidad de Santiago de Compostela, Santiago de Compostela, Spain\\
$ ^{38}$European Organization for Nuclear Research (CERN), Geneva, Switzerland\\
% ...
$ ^{41}$Nikhef National Institute for Subatomic Physics, Amsterdam, The Netherlands\\
% ...
$ ^{46}$H.H. Wills Physics Laboratory, University of Bristol, Bristol, United Kingdom\\
% ...
$ ^{51}$School of Physics and Astronomy, University of Glasgow, Glasgow, United Kingdom\\
$ ^{52}$Oliver Lodge Laboratory, University of Liverpool, Liverpool, United Kingdom\\
% ...
}
\end{flushleft}


\end{document}